\documentclass[prb,twocolumn,showpacs,preprintnumbers,amsmath,amssymb]{revtex4}
\usepackage{graphics}
\usepackage{dcolumn}

\begin{document}

\title{Smooth relativistic Hartree-Fock pseudopotentials for 
 H to Ba and Lu to Hg}

\author{J. R. Trail}
\email{jrt32@cam.ac.uk}
\affiliation{TCM Group, Cavendish Laboratory, University of Cambridge,
 Madingley Road, Cambridge, CB3 0HE, UK}
\author{R. J. Needs}
\affiliation{TCM Group, Cavendish Laboratory, University of Cambridge,
 Madingley Road, Cambridge, CB3 0HE, UK}

\date{October, 2004}

\begin{abstract}
We report smooth relativistic Hartree-Fock pseudopotentials (also
known as averaged relativistic effective potentials or AREPs) and
spin-orbit operators for the atoms H to Ba and Lu to Hg.  We remove
the unphysical extremely non-local behaviour resulting from the
exchange interaction in a controlled manner, and represent the
resulting pseudopotentials in an analytic form suitable for use within
standard quantum chemistry codes.  These pseudopotentials are suitable
for use within Hartree-Fock and correlated wave function methods,
including diffusion quantum Monte Carlo calculations.
\end{abstract}



\maketitle

Pseudopotentials or effective core potentials (ECPs) are commonly used
within electronic structure calculations to replace the chemically
inert core electrons. The influence of the core on the valence
electrons is then described by an angular-momentum-dependent effective
potential, leading to greatly improved computational efficiency in
\textit{ab initio} calculations for heavy atoms.  The use of
pseudopotentials is well established within Hartree-Fock (HF) and
Density Functional Theory (DFT), and in correlated wave function
calculations.

Our main interest is in diffusion quantum Monte Carlo (DMC)
calculations.\cite{ceperley80,foulkes_review} This technique provides
an accurate solution of the interacting electron problem for which the
computational effort scales with the number of electrons, $N$, as
approximately $N^3$, which is better than other correlated wave
function approaches. Unfortunately, scaling with atomic number, $Z$,
is approximately~\cite{ceperley86,hammond94} $Z^{5-6.5}$.  The use of
a pseudopotential reduces the effective value of $Z$, making DMC
calculations feasible for heavy atoms.

There is evidence that HF pseudopotentials give better results within
DMC than DFT pseudopotentials.\cite{greeff98} It appears that the
complete neglect of core-valence correlation within HF theory leads to
better pseudopotentials than the description of core-valence
correlation provided by DFT.  Moreover, core-valence correlation can
be included within correlated wave function calculations performed
with HF pseudopotentials by using core polarization
potentials.\cite{muller84,shirley93,lee03} Core polarization
potentials mimic the effects of dynamical polarization of the core by
the valence electrons, as well as static polarization effects due to
the other ions.  We would therefore like to use HF pseudopotentials in
our DMC calculations, preferably constructed from Dirac-Fock (DF)
theory in order to include the relativistic effects which are
significant for heavy atoms.

Standard quantum chemistry packages are convenient for generating the
``guiding wave functions'' required in DMC calculations. We would
therefore like our pseudopotentials to be available in the standard
parameterized form of a sum of Gaussian functions multiplied by powers
of the electron-nucleus separation.

Extensive sets of parameterized pseudopotentials are available in the
literature, but they have generally been constructed with different
goals to ours.  Relativistic pseudopotentials~\cite{kleinman80}
generated within DFT and the local density approximation have been
available for some time~\cite{bachelet82} but, as mentioned above, it
appears that HF pseudopotentials are superior for our purposes.

Pseudopotentials defined within HF theory have been published by Hay
and Wadt~\cite{hay-wadt} for much of the periodic table, and
Christiansen and coworkers~\cite{christiansen79,christiansen_web} have
generated similar HF pseudopotentials from DF atomic
calculations, thereby including relativistic effects.  ``Energy
consistent'' HF pseudopotentials including relativistic effects have
also been developed by the Stuttgart and Bonn groups~\cite{dolg87},
and are also publicly available.\cite{dolg_web} There is considerable
freedom in constructing pseudopotentials, and many of those available
in the quantum chemistry literature diverge at the nucleus, normally
as $1/r^2$ or $1/r$.  This singular behaviour leads to large
``time-step'' errors and even instabilities in DMC
calculations.\cite{greeff98} Moreover, we cannot imagine that this
singular behaviour is advantageous in quantum chemistry methods, as it
leads to pseudo wave functions with behaviour at small $r$ which
cannot be described within a Gaussian basis set.  All of the HF
pseudopotentials discussed above possess such singularities at the
nuclear site.

Greeff and Lester~\cite{greeff98} and Ovcharenko \textit{et
al.}~\cite{ovcharenko01} have generated non-singular HF
pseudopotentials for the atoms B-Ne and Al-Ar, which are available in
a suitable parameterized form.  These are constructed explicitly for
use in DMC calculations, and satisfy some of our criteria.
Unfortunately these pseudopotentials cover only a small part of the
periodic table, do not have $d$ angular momentum channels, do not
include relativistic effects, and are not particularly smooth even
though they are non-singular at the origin.

In this paper we report the generation of a library of accurate HF
pseudopotentials which are non-singular at the origin, include
relativistic effects, are parameterized in a form appropriate for
quantum chemistry packages, and are smooth to aid transferability.  We
also require our pseudopotentials to be well-localized (by which we
mean non-local only in a small region around the nucleus), as the
evaluation of the non-local energy within DMC is expensive.

We also seek to quantify the errors present in our pseudopotentials.
Transferability errors are expected to be the most significant, but
errors due to other sources are also assessed.  These errors result
from the approximations required to construct HF pseudopotentials from
all-electron (AE) DF data, from the removal of the unphysical long
ranged non-local behaviour discussed by Trail and Needs~\cite{trail05}, 
and from the parameterisation of the pseudopotentials.

The rest of this paper is organized as follows. In section
\ref{sec:def_pp} we describe the form of the relativistic
pseudopotentials obtained from AE DF orbitals by inversion of the HF
equations.  In section \ref{sec:gen} we describe the particular method
we use to generate pseudopotentials.  The unphysical non-local tails
of these pseudopotentials are removed and we parameterize them
accurately.
In section \ref{sec:test} we present and analyse pseudopotentials
generated for the atoms H-Ba and Lu-Hg, and present the results of
test calculations for a selection of these atoms, making comparisons
with AE DF results and with results obtained using the
pseudopotentials of Christiansen \emph{et al.} We also generate
parameterized spin-orbit (SO) pseudopotentials and compare results of
calculations using these with AE DF results, and with results obtained
using the SO pseudopotentials of Christiansen \emph{et al.}  We draw
our conclusions in section \ref{sec:conc}.

Atomic units are used throughout, unless otherwise indicated.

\section{Form of the pseudopotentials}
\label{sec:def_pp}

The coupled radial DF equations~\cite{lee77,fischer97} are
\begin{eqnarray}
\label{eq:1.1a}
\frac{d G_i}{dr} + \frac{k_i}{r}G_i -
  \left[\frac{2}{\alpha} -\alpha V^{eff,F}_{i}\right]F_{i} 
  &=& \alpha\epsilon_{i}F_{i} \; \\
\label{eq:1.1b}
\frac{d F_{i}}{dr} - \frac{k_i}{r}F_{i}-
  \alpha V^{eff,G}_{i} G_{i} 
  &=& -\alpha\epsilon_{i}G_{i} \;,
\end{eqnarray}
where $\alpha=1/c$ is the fine structure constant, $i$ is the state
index, and $G_{i}$ and $F_{i}$ are the radial components of the major
and minor (or large and small) parts of the Dirac orbitals.  The $i$
index is unique to each orbital, and hence the non-zero integer
quantum number $k_i$ can be associated with each $i$,
\begin{equation}
k_i =
  \left\{
  \begin{array}{ll}
    l_i = j_i+\frac{1}{2}       & k_i>0 \\
    -(l_i+1)=-(j_i+\frac{1}{2}) & k_i<0. 
  \end{array}
  \right.
\label{eq:1.2}
\end{equation}
The orbital dependent effective potentials, $V^{eff,G}_{i}$ and
$V^{eff,F}_{i}$, are functionals of the set $\{G,F\}$.

In what follows we also consider the standard approximation to
Eqs.~(\ref{eq:1.1a},\ref{eq:1.1b}) for valence
orbitals.\cite{kleinman80,bachelet82} Taking $\epsilon_{i}$ and
$V^{eff,F}_{i}$ to be small in Eqs.~(\ref{eq:1.1a},\ref{eq:1.1b}), as
is the case for the valence orbitals outside of the core region (this
can be taken as one of the criteria for the core/valence partition),
leads to
\begin{eqnarray}
\label{eq:1.3a}
\left[ - \frac{1}{2}\frac{d^2}{dr^2} 
       + \frac{k_i(k_i+1)}{2r^2} 
       + V^{eff,G}_{i} \right] G_{i}
&=& \epsilon_{i}G_{i} \;, \\
\label{eq:1.3b}
\frac{\alpha}{2}\left(
      \frac{d G_{i}}{dr} + \frac{k_i}{r}G_{k_i}
      \right) &=& F_{i} \;.
\end{eqnarray}
Equations~(\ref{eq:1.3a}) and (\ref{eq:1.3b}) are equivalent to
Eqs.~(\ref{eq:1.1a},\ref{eq:1.1b}) up to, but not including, terms of
order $\alpha^2$, and it should be noted that a self-consistent
eigenstate of Eqs.~(\ref{eq:1.1a},\ref{eq:1.1b}) does not satisfy
Eqs.~(\ref{eq:1.3a},\ref{eq:1.3b}) exactly in any region of space.
The effective potential in Eqs.~(\ref{eq:1.1b}) and~(\ref{eq:1.3a}) 
is given by
\begin{equation}
V^{eff,G}_i =
- \frac{Z}{r}
+ V_h[ \rho ]
+ \frac{ \hat{V}^G_x[ \{ G,F \},k_i ] G_i}{ G_i },
\label{eq:1.4}
\end{equation}
where the first term arises from the nuclear charge and the second is
the Hartree potential due to the total electron density, $\rho$.  The
third term is the effective exchange potential acting on the major
part of the radial Dirac orbital.  Both the Hartree and exchange terms
are functionals of $\{ G,F \}$, and the effective exchange potential
is different for each orbital (the cancelling self-interaction has
been included in the exchange and Hartree terms).

To generate a non-relativistic HF pseudopotential from the AE DF
solutions of Eqs.~(\ref{eq:1.1a},\ref{eq:1.1b}) we follow the
procedure described by Kleinman.\cite{kleinman80} The AE DF atomic
solutions are partitioned into core states whose influence is
represented by the pseudopotential, and valence states which are
represented by the pseudo-orbitals.  We then construct
non-relativistic pseudo-orbitals which preserve desired properties of
the original AE DF valence orbitals.

The pseudo-orbitals are chosen to satisfy the Schr\"odinger-like
equation
\begin{equation}
\left[ - \frac{1}{2}\frac{d^2}{dr^2} 
       + \frac{k_i(k_i+1)}{2r^2} 
       + {V}^{eff}_{i} \right] {\phi}_i
= \epsilon_{i} {\phi}_i \;,
\label{eq:1.5}
\end{equation}
where the pseudo-orbital, ${\phi}_i$, is a scalar orbital and the
corresponding eigenvalue $\epsilon_i$ is equal to that of
Eqs.~(\ref{eq:1.1a},\ref{eq:1.1b}) for the AE orbital $i$.  We may
relate Eq.~(\ref{eq:1.5}) to the Dirac-Fock equation by taking the
limit $\alpha \rightarrow 0$ in Eqs.~(\ref{eq:1.3a},\ref{eq:1.3b}),
which leads to a scaler Schr\"odinger equation.

The effective potential, ${V}^{eff}_i$ for state $i$ is then given by 
\begin{equation}
{V}^{eff}_i =
  {V}_i(r)
+ V_h[ {\rho^{pseudo}} ]
+ \frac{ \hat{V}^G_x[ \{ {\phi},0 \},k_i ] {\phi}_i}{ {\phi}_i },
\label{eq:1.6}
\end{equation}
where $\rho^{pseudo}$ is the charge density obtained from the occupied
pseudo states.  This defines the pseudopotential for orbital $i$,
${V}_i(r)$, in terms of the pseudo-orbital ${\phi}_i$.  Given
${\phi}_i$ and $\epsilon_i$, ${V}_i(r)$ may be obtained by direct
inversion of Eqs.~(\ref{eq:1.5}) and (\ref{eq:1.6}).

In the non-relativistic case we would construct valence
pseudo-orbitals, ${\phi}_i$, which are equal to the equivalent AE
orbitals outside of a `core radius', $r_{ci}$.  In the relativistic
case this procedure must be modified, since the DF orbitals consist of
two components.  We could simply take ${\phi}_i = G_{i}$ for
$r>r_{ci}$,
but this is unsatisfactory because it implies that charge has been
removed from the system.  Instead we take the form
\begin{equation}
{\phi}_i(r) =
  \left\{
  \begin{array}{ll}
    f_i(r)                                                 & r <    r_{ci} \\
    \left[ G_{i}(r)^2 + F_{i}(r)^2  \right]^{\frac{1}{2}}  & r \geq r_{ci}\;,
  \end{array}
  \right.
\label{eq:1.7}
\end{equation}
which preserves the valence charge density outside of the core region.
Inside $r_{ci}$ the orbitals are given by the, as yet, unspecified
function $f_i$. This function is generally chosen such that ${\phi}_i$
is node-less, smooth at $r_{ci}$ to a certain order of
differentiation, and that it satisfies the norm-conservation
condition, that is, the total charge inside of $r_{ci}$ is the same as
for the AE orbital.

The pseudopotentials ${V}_i$ are not quite appropriate for use in
non-relativistic calculations as they provide different
pseudopotentials for the same $l$ values.  However, the valence states
chosen to construct the pseudopotential have unique $k$ numbers, and
therefore the index $k$ is interchangeable with the pseudo-atom
orbital index, $i$.  Taking the average of the pseudopotentials with
the same $l$ quantum numbers weighted by the different
$j$~degeneracies, we obtain the averaged relativistic effective
potential (AREP)~\cite{kleinman80,bachelet82b},
\begin{equation}
V^{p}_{l} = \frac{1}{2l+1}\left[
l{V}_{k_i=l} + (l+1){V}_{k_i=-l-1}
\right] \; .
\label{eq:1.8}
\end{equation}
This includes all of the Dirac relativistic effects except the SO
coupling.  The SO operator, $\hat{V}^{so}_{l}$, may be expressed in
terms of a SO pseudopotential, $V^{so}_{l}$, and angular momentum
operators,
\begin{eqnarray}
\hat{V}^{so}_{l} &=& V^{so}_{l} \hat{\mathbf{L}}\cdot\hat{\mathbf{S}} \\
                 &=& \frac{2}{2l+2} \left[
{V}_{k_i=-l-1} - {V}_{k_i=l}
\right] \hat{\mathbf{L}}\cdot\hat{\mathbf{S}} \; .
\label{eq:1.9}
\end{eqnarray}

To use the pseudopotential in a calculation for a molecule or solid it
is expressed in terms of projection operators, and separated into
local and non-local parts,
\begin{eqnarray}
\hat{V}_{pseudo}&=&V^p_{local}(r) \nonumber \\*
   & & + \sum_{l}^{l_{max}} \sum_{m=-l }^{l}
  | Y_{lm} \rangle ( V^p_{l}(r) - V^p_{local}(r))
   \langle Y_{lm} |. \nonumber \\*
\label{eq:1.10}
\end{eqnarray}
Orbitals with $l>l_{max}$ feel the local potential, $V^p_{local}$.

\section{Generation of the pseudopotentials}
\label{sec:gen}

We take the core states to be the `noble core' (for example, for Si we
take $1s2s2p$ as core states and $3s3p$ as valence states) or noble
core plus a filled $d$ shell.  We generate pseudopotentials for the
$s$,$p$ and $d$ channels.
We use the atomic ground state configurations to obtain channels for
which the corresponding AE valence orbitals are bound.  For channels
which have no corresponding AE valence orbitals in the ground state we
use the excited state configurations of Bachelet \textit{et
al.}~\cite{bachelet82}, which provide an appropriate bound state.
Fischer's interpretation of fractional occupation numbers within HF
and DF theory~\cite{fischer_hf1} is used. In most cases we choose the
core radii, $r_{ci}$, to be 0.9 of the distance from the outermost
node to the outermost maximum, although for some cases we choose
smaller radii.

To define our pseudo-orbitals we use the Troullier-Martins
scheme~\cite{troullier91} where the pseudo-orbitals within the core
region are given by
\begin{equation}
f_i(r) = r^{l_i+1}\exp\left[\sum_{m=0}^6 c_{2m} r^{2m}\right] ,
\label{eq:2.1}
\end{equation}
and the coefficients $c_{2m}$ are to be determined.  The factor
$r^{l_i+1}$ ensures that no $1/r^2$ singularity occurs in the
effective potential (and resulting pseudopotential), and $c_1=0$
ensures that no $1/r$ singularity is present.  All of the $c_{2m+1}$
terms have been excluded to prevent the appearance of a cusp of any
order in the pseudopotentials at the origin, so improving the
asymptotic behaviour of the pseudo-orbitals and pseudopotentials in
momentum space.
The seven coefficients in Eq.~(\ref{eq:2.1}) are determined by the
conditions of:
\begin{enumerate}
\item norm-conservation within $r_{ci}$,
\begin{equation}
\int_0^{r_{ci}} {\phi}_i(r)^2                          dr =
\int_0^{r_{ci}} \left[ G_{i}(r)^2 + F_{i}(r)^2 \right] dr
\; ;
\label{eq:2.2}
\end{equation}
\item continuity of ${\phi}_i(r)$ and its first four derivatives at
$r_{ci}$;
\item zero curvature of the screened potential at the origin,
\begin{equation}
\left. \frac{d^2 {V}^{eff}_{i}}{dr^2}\right|_{r=0} = 0 \;.
\label{eq:2.3}
\end{equation}
\end{enumerate}
In the core region we obtain ${V}^{eff}_{i}$ by inversion of
Eq.~(\ref{eq:1.5}), and obtain ${V}_i$ by `unscreening'
${V}^{eff}_{i}$ using the pseudo-orbitals.

Outside of the core region we could in principle apply the same
inversion procedure, but this would involve taking numerical
differences of small quantities, which is prone to errors.  Instead we
note that the eigenvalues of the AE DF and pseudo valence states are
equal, and that we expect Eqs.~(\ref{eq:1.3a},\ref{eq:1.3b}) to be an
extremely good approximation for valence states in the region outside
of the core.  We may then use Eqs.~(\ref{eq:1.3a}), (\ref{eq:1.4}),
(\ref{eq:1.5}), and (\ref{eq:1.6}) to obtain
\begin{eqnarray}
{V}_i&=& - \frac{Z}{r} + V_h[ \rho - \rho^{pseudo} ] + \nonumber \\* &
& \frac{ \hat{V}^{G}_x [ \{ G ,F \}, k_i ] G_{i} }{ G_{i} } - \frac{
\hat{V}^{G}_x [ \{ {\phi},0 \}, k_i ] {\phi}_{i} }{ \phi_{i} } -
\nonumber \\* & & \frac{1}{2}\left( \frac{G''}{G} -
\frac{{\phi}''}{{\phi}} \right)\;,
\label{eq:2.4}
\end{eqnarray}
for $r>r_{cl}$.  The final term in this expression arises from the
kinetic energy, and we neglect this, which introduces an error of
order $\alpha^2$.  This approximation may seem unnecessary, but errors
of this order are already present if we assume that the pseudo-valence
electrons may be treated as scalar relativistic, and it has the
advantage that the valence charge density outside of the core region
is preserved.  We also found that this error, when detectable, is
smaller than that due to the fact that Eq.~(\ref{eq:1.3a}) is already
an approximation, even outside of the core region.  Some
authors\cite{bachelet82} have chosen ${\phi}_i=\lambda G_i$ outside of
the core, so that the kinetic terms in Eq.~(\ref{eq:2.4}) are zero.
The $\lambda$ constant may be chosen to ensure that ${\phi}_i$ is
normalized.  This would cause a negligible (in most cases
undetectable) change in the pseudopotentials generated here.

All DF and HF calculations were carried out using fine radial grids
and therefore do not suffer from basis set errors.  We used the DF
code of Ankudinov \textit{et al.}~\cite{ankudinov} for the AE DF
calculations, the HF code of Fischer~\cite{fischer_hf2} for the AE HF
calculations, and the pseudopotential HF calculations were performed
using our own code. Grid and convergence parameters were chosen so as
to achieve $10$-digit accuracy in the total energy, and the ground
state energies were checked against those of Visscher and
Dyall~\cite{visscher97}, and Librelon and Jorge.\cite{librelon03} The
above procedure provides us with relativistic screened
pseudopotentials for each of the valence states considered and, from
Eq.~(\ref{eq:2.4}), the ${V}_{i}$.  We then use Eq.~(\ref{eq:1.8}) and
(\ref{eq:1.9}) to obtain the AREP and SO potentials.

\subsection{Non-local asymptotic behaviour}

In a previous paper~\cite{trail05} we examined pseudopotentials
constructed within HF theory using this inversion procedure, and found
them to be non-local over all space, the deviation from the ionic
Coulomb potential remaining finite as $r\rightarrow \infty$.  As the
presence of this `extreme non-locality' is a consequence of the
non-locality of the exchange interaction in HF theory, and the DF
exchange interaction is of essentially the same form, this effect also
occurs for the pseudopotentials defined here.

The presence of this long-ranged non-locality leads to a loss of
transferability and problems in defining a total energy for extended
systems.  Here we remove the `extreme non-locality' using the method
presented in our previous paper~\cite{trail05}, which we briefly
summarize here.

In order to generate a new version of the pseudopotential,
$V^{p,loc}_l(r)$, which is non-local only close to the atomic core,
the original pseudopotential is transformed using
\begin{equation}
V^{p,loc}_l(r) =
  \left\{
  \begin{array}{ll}
    \gamma_l(r) + V^p_l(r)               & r < r_c \vspace{+0.2cm} \\
    e^{-\eta(r-r_c)^2} \times            &         \\
    ~  \left(
       \gamma_l(r) + V^p_l(r)
       - V_h[ \rho_{core} ] + \frac{Z}{r}
       \right)                           & \\
    ~  +V_h[ \rho_{core}] - \frac{Z}{r}  & r \geq r_c \;,
  \end{array}
  \right.
\label{eq:2.5}
\end{equation}
where $V_h[ \rho_{core} ] - Z/r$ is the ionic potential and
$\eta^{-\frac{1}{2}} = \max_l\left[r_{cl}\right]/16$ is a parameter
which specifies the length scale over which the transformed
pseudopotential becomes local outside of the core region.  The
function $\gamma_l$ is given by
\begin{equation}
\gamma_l(r) =
  \left\{
  \begin{array}{ll}
    q_l + p_l r^4 \left(1 - \frac{2}{3r_c^2} r^2 \right)  & r <    r_c \\
    q_l + p_l \frac{r_c^4}{3}                             & r \geq r_c \;,
  \end{array}
  \right.
\label{eq:2.6}
\end{equation}
where the parameters $q_l$ and $p_l$ are chosen such that the HF
eigenvalues of the original pseudopotential are preserved, and that
the logarithmic derivatives of the original pseudo-orbitals 
at $r_{ci}$ are preserved to high accuracy.~\cite{trail05}
Although the norm of the original
pseudopotential is not exactly preserved this was not found to
significantly affect transferability.

\subsection{Expansion in a Gaussian basis}

For applications within quantum chemistry codes it is normally
necessary to have the pseudopotentials available as an expansion in
Gaussian functions.  The pseudopotentials described so far are
tabulated on a radial grid, and in this section we develop a Gaussian
fitting procedure which is both accurate and reliable.

Our aim is to find an expansion of the localized AREP $V^{p,loc}_{l}$
in a standard Gaussian form,
\begin{equation}
\tilde{V}^{p}_{l} = \sum_{q}^{q_{max}} A_{ql} r^{n_{ql}} e^{-a_{ql}r^2} =
  \left\{
  \begin{array}{rr}
   Z_{val}/r        + \tilde{V}^{p}_{local} & l  =   {\rm local}\\
   \tilde{V}^{p}_l  - \tilde{V}^{p}_{local} & l \neq {\rm local}
  \end{array}
  \right. \;,
\label{eq:2.7}
\end{equation}
where $Z_{val}$ is the number of valence electrons in the neutral
atom.  A set of $n_{ql}$ is chosen such that the pseudopotential can
be represented accurately by the expansion, with no singularity
present at the origin.  Once we have this expansion it may be applied
in the same projector form as Eq.~(\ref{eq:1.10}).  It is important to
note that we expand the \textit{localized} tabulated pseudopotential
in terms of Gaussian functions.  This is desirable since the removal
of the long-ranged non-local tail significantly improves the accuracy
and stability of any fitting procedure using localized basis
functions, such as the Gaussian expansion.

In order arrive at an algorithm to generate our expansion we must
decide what properties a `good' Gaussian expansion must have.  We
require a good expansion to reproduce both the eigenvalues and
orbitals of the tabulated pseudopotential to high accuracy, and so
reproduce the scattering properties of the tabulated pseudopotential.
We require a good expansion to be close to the tabulated
pseudopotentials in the least-squares sense, because such deviations
tend to make the fitted ones less smooth and reduce the
transferability.  We also require the parameterized pseudopotentials
to be non-local only in a region near the core, although this
criterion is partly included in the previous two, together with the
use of the localized tabulated pseudopotential.  These are stringent
requirements and we therefore require a rather larger expansion than
has been used in earlier work.  Finally we require an expansion which
is compatible with quantum chemistry codes, and we have successfully
tested our parameterized form in both the
\textsc{crystal}~\cite{crystal_98} and
\textsc{gaussian}~\cite{gaussian_03} codes.

We chose the local part of the pseudopotential to be $l=2$ for all
atoms, and to obtain an accurate (in the least squares sense)
representation of the pseudopotential we chose $q_{max}=8$, with
$n_{ql}=-1,0,1,2$ for the local part and $n_{ql}=0,1,2$ for the
non-local parts.  This gives $16$ parameters for each channel.  It is
helpful to reduce this number by imposing constraints on the
functional form of the pseudopotential.  We require that
$\tilde{V}^{p}_{l}$ is finite and has zero derivative at the nucleus,
as did Greeff and Lester.\cite{greeff98} In addition we require the
second derivative of $\tilde{V}^{p}_l$ to be zero at the nucleus and
that $\tilde{V}^{p}_l(0) = {V}^{p,loc}_l(0)$.  The latter two
conditions make $\tilde{V}^{p}_l$ smoother because they force it to be
closer to the original tabulated pseudopotential.  These conditions
are imposed by reducing the number of free $A_{ql}$ parameters.

Since Eq.~(\ref{eq:2.7}) is linear in the $\{A_{ql}\}$ it is also
possible to use Singular Value Decomposition (SVD) to define the
remaining $A_{ql}$ parameters as those which give the minimum least
squares deviation from the original tabulated pseudopotential.  

A further (nonlinear) least squares minimization with respect to the
remaining free parameters $\{a_{ql}\}$ provides a `good fit' to the
tabulated pseudopotential, but does not provide an expansion which
accurately reproduces the eigenstates of the original pseudopotential.
We take the $\{a_{ql}\}$ parameters obtained from this least squares
procedure as a starting point for a second stage of optimization.
Here we employ a generalization of an algorithm developed by Barthelat
\textit{et al.}\cite{barthelat77}

To progress further we perform a HF atomic calculation with $LS$
coupling and using the tabulated pseudopotential $V^{p,loc}_l$,
yielding the pseudo-states $\{{\phi}^{p}_l$,$\epsilon^{p}_l\}$.  For a
given set of parameters for each channel, we also perform a similar HF
atomic calculation using the parameterized pseudopotential
$\tilde{V}^{p}_l$, to give the pseudo-states
$\{\tilde{\phi}^{p}_l$,$\tilde{\epsilon}^{p}_l\}$.  We then define a
functional of the pseudo-states of both the tabulated and
parameterized pseudopotentials,
\begin{equation}
\Sigma=\sum_l \langle {\phi}^{p}_l 
               | \hat{O}^2_l | 
                      {\phi}^{p}_l \rangle \;,
\label{eq:2.8}
\end{equation}
where
\begin{equation}
\hat{O}_l =
\tilde{\epsilon}^{p}_l | 
   \tilde{\phi}_l^{p} \rangle \langle 
   \tilde{\phi}_l^{p} | -
\epsilon^{p}_l | 
   {\phi}_l^{p} \rangle \langle  
   {\phi}_l^{p} | \;,
\label{eq:2.9}
\end{equation}
which is a function of the $\{a_{ql}\}$ for all $l$ which correspond
to valence states in the atomic configuration considered.  We vary the
available $a_{ql}$ of the parameterized pseudopotential to search for
a minimum of $\Sigma$ where the overlap $\langle \tilde{\phi}_l^{p} |
{\phi}_l^{p} \rangle$ is maximized and the error in the eigenvalue
$|\tilde{\epsilon}^{p}_l - \epsilon^{p}_l|$ is minimized separately
for each available $l$.  In the original formulation of Barthelat
\textit{et al.}~\cite{barthelat77} no sum over states was present in
Eq.~(\ref{eq:2.8}), and each channel was optimized separately.  We
found that optimizing each channel separately did not allow
convergence to be achieved to a high enough accuracy for the majority
of atoms, whereas optimization over all states provided reliable
convergence.

A standard quasi-Newton minimization algorithm was used in the
optimization, and we found that the stability and efficiency could be
improved by introducing
\begin{equation}
a'_{ql}=\ln a_{ql}\;,
\label{eq:2.10}
\end{equation}
and minimizing with respect to the $\{ a'_{ql} \}$.

We considered the optimization to be successful when the pseudo-states
of the parameterized pseudopotentials satisfied the conditions
\begin{equation}
  \begin{array}{rcl}
  1-\langle \tilde{\phi}^{p}_l| {\phi}^{p}_l \rangle &<& 10^{-6} \;, \\
  |\tilde{\epsilon}^{p}_l-\epsilon^{p}_l|            &<& 10^{-5} \; {\rm a.u.} \;,
  \end{array}
\label{eq:2.11}
\end{equation}
for the $s$ and $p$ block atoms, and
\begin{equation}
  \begin{array}{rcl}
  1-\langle \tilde{\phi}^{p}_l| {\phi}^{p}_l \rangle &<& 10^{-5} \;, \\
  |\tilde{\epsilon}^{p}_l-\epsilon^{p}_l|            &<& 10^{-4} \; {\rm a.u.} \;,
  \end{array}
\label{eq:2.12}
\end{equation}
for the transition metal atoms.

On implementation it became apparent that, for most atoms, the final
values of the parameters are surprisingly sensitive to their initial
values, suggesting that $\Sigma$ has many minima.
We also found that, without the SVD definition of the $A_{ql}$
parameters, the rescaling of Eq.~(\ref{eq:2.10}) or the concomitant
optimization implied by the sum in Eq.~(\ref{eq:2.8}), this procedure
did not converge for many atoms, suggesting that these aspects of the
algorithm are necessary to provide a functional with minima which are
distinct and well defined.  The presence of many locally optimum
parameter sets is also an advantage, as when a set of optimized
parameters does not satisfy our criteria a new contender may easily be
obtained by making a small change to the initial parameter set.

\begin{figure}
\includegraphics{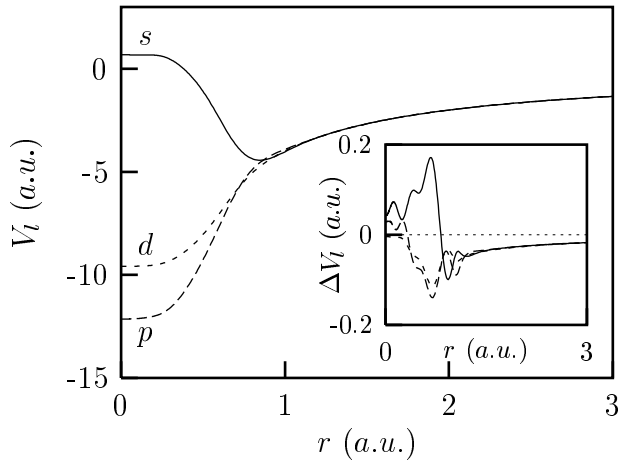}
\caption{\label{fig:1} Our parameterized pseudopotential for C. On 
the scale shown it is indistinguishable from both the original tabulated 
TM pseudopotential and its localized version.  The inset shows the
difference between the parameterized and original TM
pseudopotentials.}
\end{figure}

As an example we consider C, for which our parameterized
pseudopotential is shown in Fig.~\ref{fig:1}.  The $s$ and $p$
channels were obtained from the neutral ground state, while the $d$
channel was obtained from an excited ionic state
($1s^{2}2s^{0.75}2p^{1.00}3d^{0.25}$) taken from Bachelet \textit{et
al.}\cite{bachelet82} Inset in the same figure is the difference
between the parameterized pseudopotential and the original TM
(extremely non-local) pseudopotential.  This figure demonstrates that
our parameterisation accurately reproduces the original
pseudopotential, with the greatest difference occurring at small $r$.
Most of this difference is not due to the fitting procedure - it is
due to the change in the pseudopotential required to enforce locality
outside of a small radius (in this case $r_{loc}=1.2$ a.u.)  while
preserving the eigenvalues and logarithmic derivatives of the original
pseudo-states.

\begin{figure}
\includegraphics{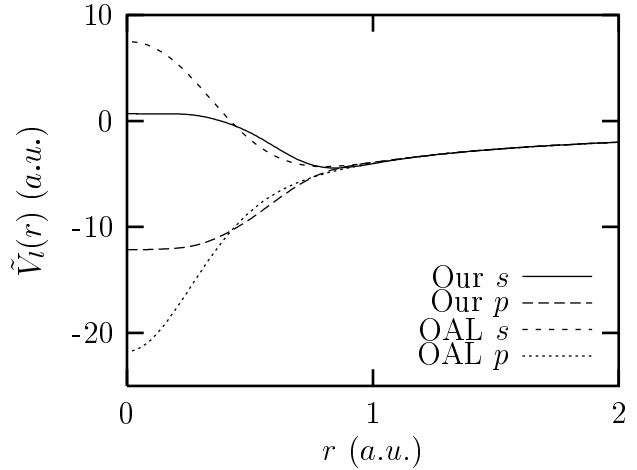}
\caption{\label{fig:2} Comparison of the $s$ and $p$ channels of the 
parameterized C pseudopotentials generated within this paper (Our) and 
by Ovcharenko \emph{et al.}~\cite{ovcharenko01} (OAL).  (The OAL 
pseudopotential does not have a $d$ channel for us to compare with.)}
\end{figure}

In Fig.~\ref{fig:2} the $s$ and $p$ components of our C
pseudopotential are compared with the non-singular (non-relativistic)
pseudopotential published by Ovcharenko \textit{et
al.}\cite{ovcharenko01} Fig.~\ref{fig:2} shows that our C
pseudopotential is smoother than that of Ovcharenko \textit{et al.},
and we found this to be the case for all of their pseudopotentials.
This is probably due to the extra constraints we apply in our
construction of the pseudo-orbitals.

\begin{figure}
\includegraphics{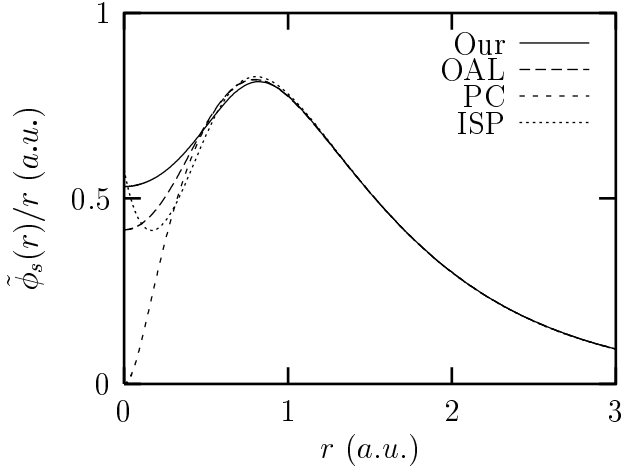}
\caption{\label{fig:3} Comparison of the $s$ pseudo-orbital from the 
neutral ground state of our parameterized C pseudopotential with those 
generated from the parameterized pseudopotentials of Ovcharenko \textit{et
al.}~\cite{ovcharenko01} (OAL), Pacios and
Christiansen~\cite{christiansen79,christiansen_web} (PC) and Igel-Mann
\textit{et al.}\cite{igel-mann}}
\end{figure}

Fig.~\ref{fig:3} shows the $s$ pseudo-orbitals resulting from our C
pseudopotential, together with those from the pseudopotentials of
Ovcharenko \textit{et al.}~\cite{ovcharenko01} (OAL), Christiansen
\emph{et al.}~\cite{christiansen_web} (PC), and those of the Stuttgart
group~\cite{igel-mann} (ISP).  The $+1/r^2$ divergence of the PC
pseudopotential forces the $s$ pseudo-orbital to go rapidly to zero at
the nucleus, while the $-1/r$ divergence of the ISP pseudopotential
forces the $s$ pseudo-orbital to have a cusp at the nucleus.  Our
pseudopotential gives the smoothest pseudo-orbital of those plotted.

We have generated pseudopotentials using the scheme described above
for the atoms H to Ba and Lu to Hg.  We include H and He even though
they have no core electrons because the smoothness of the resulting
pseudopotentials may make them useful in some circumstances.

To provide a quantitative measure of the errors introduced by the
various approximations used in the construction of the
pseudopotentials we investigated how well the pseudopotentials
reproduce the AE atomic results at the various levels of
approximation.
Atomic $LS$ and $jj$ coupled HF calculations were performed with the
pseudopotentials using our own code.  Here we report detailed results
for the Li, Be, C, Si, Ti, Cr, Fe, Br, Mo, Ag, and Sb atoms, which are
representative of the general accuracy achieved.  Note that we have
deliberately chosen several transition metal atoms as they generally
exhibit the largest errors.

If there were no approximations in our pseudopotential construction
procedure, then a HF pseudopotential calculation with the exchange
interaction defined using $jj$ coupling should exactly reproduce the
AE DF eigenvalues of the occupied levels of the ground state.  In
practice, however, this is not achieved, and the pseudopotential
generation method we use may be viewed as introducing four types of
error.  These are shown in Table~\ref{tab:1} together with the AE DF
eigenvalues, and the $jj$ coupled HF eigenvalues resulting from the
final parameterized pseudopotential.  For both the errors and the
eigenvalues themselves we take an average over values corresponding to
the same $l$ weighted by the different $j$-degeneracies, just as we
did to construct the AREP.

The first type of error, $\Delta\epsilon_1$, is introduced by the fact
that the pseudo-orbital is not an exact solution of
Eq.~(\ref{eq:1.3a}) outside of the core region.  The second error,
$\Delta\epsilon_2$, is introduced by the $j$-averaging of the
pseudopotential.  This error results from the exclusion of SO coupling
from the pseudopotential, and is due to the higher order effects of SO
coupling.  The third error $\Delta\epsilon_3$ is introduced by the
localization of the pseudopotential (by construction this error would
be zero for $LS$ coupling).  Finally, the fourth error,
($\Delta\epsilon_4$), is introduced by the imperfect parameterisation.
These four errors are shown in Table~\ref{tab:1} and are defined such
that their sum is the total error of the (averaged) eigenvalues
resulting from the parameterized pseudopotential.  Results are given
for the atoms listed above, together with the separate averages over
the $s$, $p$ and $d$ block atoms considered in this paper.

The total error is small for all atoms, it is negligible for the
$s$-block atoms and is largest for the heaviest transition metal
atoms.  Generally, the largest total errors are dominated by the error
due to the removal of SO coupling, this error increasing rapidly with
increasing atomic number (compare $\Delta\epsilon_2$ for Mo and Ag).
$\Delta\epsilon_1+\Delta\epsilon_3+\Delta\epsilon_4$ remains below
$10^{-4}$ a.u., with the exception of a few heavy transition metal
atoms such as Mo.  Errors due to the approximation of the Dirac
equation as a Schr\"odinger equation in the valence region remain
consistently small, and errors due to localization and
parameterisation are largest for the transition metal atoms.

In Table~\ref{tab:2} we compare the $j$-averaged eigenvalues resulting
from AE DF calculations, eigenvalues resulting from HF calculations
using our soft parameterized pseudopotentials, and eigenvalues
resulting from the singular PC pseudopotentials.  For all comparisons
with PC pseudopotentials we choose those constructed using the same
valence/core partition as our own.  The $j$-averaged valence
eigenvalues of the original AE DF atom are well reproduced by both the
soft and singular pseudopotentials, with our soft pseudopotentials
performing slightly better.  We conclude that our pseudopotentials
have been parameterized successfully.

\section{Testing of the pseudopotentials}
\label{sec:test}

The analysis in the previous section demonstrates that our
pseudopotentials have been parameterized accurately.  Of course this
does not demonstrate their transferability, and it is this issue which
is addressed in this section.

Although designed to reproduce $jj$ coupling results, our
pseudopotentials are intended for use within $LS$ coupling
calculations, and therefore we have calculated HF excitation energies
and ionization energies within $LS$ coupling for both our
pseudopotentials and those of Christiansen \textit{et
al.}~\cite{christiansen_web} (PC).  Both of these types of
pseudopotential contain relativistic effects at the DF level, and
therefore we compare these results with AE HF results in which scalar
relativistic effects are incorporated perturbatively.  The
calculations were performed using Fischer's code with the Breit-Pauli
scalar relativistic correction added to the total energy, correct to
order $\alpha^2$.  This order of correction would not be particularly
accurate for AE calculations on the heavier atoms, but
these errors arise largely from the deep core electrons and do not
contribute significantly to the energy differences between the
low-lying excited states considered here.

In Table~\ref{tab:3} we compare excitation energies for the same atoms
considered in Tables~\ref{tab:1} and \ref{tab:2}. Excitation energies
obtained for the AE atoms are compared with those from our soft AREP
parameterized pseudopotentials and the singular PC pseudopotentials.
We find that our pseudopotentials reproduce the excitation energies to
the same accuracy as those of PC.

We also consider the errors for our full set of pseudopotentials in
reproducing the AE DF excitation energies.
It is apparent that the excitation energies are reproduced most
accurately for the $s$-block atoms with an average absolute error of
$0.00084$~a.u.~(0.022~eV), that they are less well reproduced for
$p$-block atoms with an average absolute error of
$0.00344$~a.u.~(0.090~eV), and that the largest errors occur for the
transition metal atoms, with an average absolute error of
$0.01204$~a.u.~(0.32~eV).  The large error value for the transition
metals is mainly due to the $5d$ transitions - the average absolute
errors are $0.00687$~a.u.~(0.18~eV), $0.00754$~a.u.~(0.20~eV), and
$0.02116$~a.u.~(0.55~eV) for the $3d$, $4d$, and $5d$ transitions metals
respectively.  Further examination of the results reveals that these
errors are inherent in the pseudopotentials themselves, and not due to
the errors examined in the previous section.  Our pseudopotentials and
the PC pseudopotentials also appear to produce very similar errors,
providing support for this interpretation.  For the transition metal
atoms it should be possible to reduce these errors by including the
$3s$ and 3$p$ electrons as valence states, giving ``small core''
pseudopotentials.\cite{smallcore}

In Table~\ref{tab:4} we compare the first ionization potentials
calculated for the atoms considered in the previous tables.  The
results are of similar quality to those for the excitation energies
given in Table~\ref{tab:3}, and show a similar trend that the errors
tend to increase with $Z$.  The PC pseudopotentials perform slightly
better for the ionization potentials than our soft pseudopotentials,
but the difference does not appear to be significant considering the
number of atoms studied.

Our pseudopotential generation procedure also provides a description
of SO splitting in terms of the SO pseudopotential, $V^{so}_l$ defined
in Eq.~(\ref{eq:1.9}).  We may define a Gaussian parameterisation of
this as
\begin{equation}
 \tilde{V}^{so}_{l} =
 \sum_{q}^{q_{max}} B_{ql} r^{n_{ql}} e^{-b_{ql}r^2}\;,
\end{equation}
where $\tilde{V}^{so}_{l}$ is the fitted potential and the $B_{ql}$
and $b_{ql}$ play the same role as the $A_{ql}$ and $a_{ql}$ in the
AREP.  No Coulomb potential terms are required, and no local potential
term is required since $\tilde{V}^{so}_{l}=0$ for $l>l_{max}$ is a
physically reasonable approximation.  In addition we require
$B_{ql}=0$ for $l=0$, since $V^{so}_{l}=0$ for $l=0$.

It would be possible to follow the same route of localization and
parameterisation used for the AREP. This would require $jj$ coupled HF
calculations to be used in the localization and fitting procedures of
section \ref{sec:gen} (as opposed to $LS$ coupled HF), so we could
seek a $\tilde{V}^{so}_{l}$ which reproduces the SO splitting of the
original AE DF calculation.  Here we use a simpler procedure and apply
least squares fitting with respect to $\{B_{ql},b_{ql}\}$, with the
initial $\{b_{ql}\}$ taken as the values of the equivalent AREP
parameters.

We considered the selection of atoms and ions shown in
Table~\ref{tab:5}.  The SO splitting is defined as the difference
between the energies of states where $J$ differs by one due to the
transfer of an electron between orbitals of the same $l$, but
different $j$ quantum number.  The difference in energy of the states
due to the SO interaction was obtained from AE DF calculations (not
including further relativistic corrections), and from $jj$-coupled HF
calculations with the parameterized pseudopotentials.
For the pseudopotential calculations
$\tilde{V}^{p}$ and $\tilde{V}^{so}_{l}$ are used to construct a
Relativistic Effective Potential (REP) that depends on $j$ - the
parameterized equivalent of $V_{i}$ in section \ref{sec:def_pp} - and
this REP provides the external potential.  Since the SO splitting is
not obtained perturbatively any differences between AE and AREP$+$SO
results are solely due to the REP.

We calculated energy differences between $2P_{1/2}$ and $2P_{3/2}$
states for $s$ and $p$-block atoms, and between $2D_{3/2}$ and
$2D_{5/2}$ states for $d$-block atoms. In Table~\ref{tab:5} we compare
the SO splitting energies obtained from AE DF calculations, those
obtained using our parameterisation of $\tilde{V}^{p}$ and
$\tilde{V}^{so}_{l}$, and those obtained using the (singular)
parameterized SO pseudopotentials published by Pacios and
Christiansen.\cite{pacios85,hurley86a,christiansen_web} 
Our AE results differ somewhat from those reported by Christiansen 
\textit{et al.}~\cite{ross86b}, who used the code of 
Desclaux~\cite{desclaux75}.
These discrepancies appear to be due to the finer radial grids and
higher tolerances used in our calculations.
We have also found discrepancies between our results and
Christiansen's \textit{et al.}~\cite{ross86b} using their REPs, which we
believe is due to the same effect.  We also note that our code reproduces
the accurate ground state energies of Visscher and Dyall
\cite{visscher97}, but the code of Desclaux~\cite{desclaux75} does not.

We find that in most cases our pseudopotentials reproduce the DF
splitting energies more accurately than those of PC, with an average
absolute percentage error of $4.8~\%$ compared to $8.2~\%$ for PC.
The largest percentage error is for Li ($27~\%$, compared with $34~\%$
from PC), where the splitting is small. For the rest of the atoms
considered, the errors are $< 11~\%$, compared with $< 24~\%$ for PC.

Parameter values for our pseudopotentials and SO pseudopotentials 
are provided as supplementary material\cite{epaps}, and are given 
to the same numerical precision used to generate the results in 
this paper.

\section{Conclusions}
\label{sec:conc}

We have developed smooth HF pseudopotentials which emulate the
influence of relativistic core electrons for the atoms H to Ba and Lu
to Hg.  These are the Averaged Relativistic Effective Potentials and
SO pseudopotentials of Kleinman~\cite{kleinman80} and Bachelet and
Schl\"uter.\cite{bachelet82b} We use the Troullier-Martins
scheme~\cite{troullier91} to generate smooth pseudopotentials, and
remove the unphysical extremely non-local behaviour which results from
the exact exchange in a controlled manner.\cite{trail05}

The resulting (tabulated) pseudopotentials are then represented in a
convenient analytic form suitable for use in standard quantum
chemistry codes, using a fitting procedure based on the method first
described by Barthelat \textit{et al.}\cite{barthelat77} The version
of this scheme employed here increases the efficiency and accuracy of
the fitting procedure, and preserves the smoothness of the
Troullier-Martins pseudopotentials.

An analysis of the performance of our parameterized pseudopotentials
for a number of test cases reveals that they reproduce the original AE
atomic results well in most cases, and perform well in comparison with
pseudopotentials published by previous authors which possess
singularities at the nuclei.

The pseudopotentials we have generated are finite at the nucleus and
are smoother than other HF based pseudopotentials, which should aid
the convergence of methods employing Gaussian basis sets.  These
pseudopotentials are appropriate for use within non-relativistic
theories of the valence electrons, such as HF theory or correlated wave
function methods, including DMC.

\begin{acknowledgements}
We would like to thank Dr. Y. Lee, whose Ph.D. \cite{lee02} work
provided the algorithm which has evolved into the work presented here.
Financial support was provided by the Engineering and Physical
Sciences Research Council (EPSRC), UK.
\end{acknowledgements}


\begin{table}[t]
\begin{tabular}{lcccrrrrr}
\hline \hline
\multicolumn{4}{c}{ }                      &
\multicolumn{1}{c}{$\Delta\epsilon_1$}     &
\multicolumn{1}{c}{$\Delta\epsilon_2$}     & 
\multicolumn{1}{c}{$\Delta\epsilon_3$}     &
\multicolumn{1}{c}{$\Delta\epsilon_4$}     &
\multicolumn{1}{c}{$\Delta\epsilon_{tot}$} \\
\multicolumn{1}{c}{Atom}                   &
\multicolumn{1}{c}{$l$}                    & 
\multicolumn{1}{c}{$\epsilon^{AE}_l$}      &
\multicolumn{1}{c}{$\epsilon^{jj}_l$}      &
\multicolumn{5}{c}{$\times 10^{-5}$}       \\
\hline
Li &$s$& -0.19634 & -0.19634 &  0.0 &  0.0 &  -0.0 &   0.0 &   0.0 \vspace{+0.1cm} \\
Be &$s$& -0.30932 & -0.30933 & -0.1 &  0.0 &  -0.0 &  -0.4 &  -0.4 \vspace{+0.1cm} \\
C  &$s$& -0.71212 & -0.71213 & -0.9 &  0.0 &  -0.0 &   0.4 &  -0.5 \vspace{ 0.0cm} \\
   &$p$& -0.40867 & -0.40869 &  0.3 &  0.1 &  -2.2 &  -0.5 &  -2.3 \vspace{+0.1cm} \\
Si &$s$& -0.54522 & -0.54522 & -0.2 &  0.0 &   0.1 &   0.3 &   0.2 \vspace{ 0.0cm} \\
   &$p$& -0.27967 & -0.27968 &  0.0 &  0.3 &  -1.1 &  -0.6 &  -1.3 \vspace{+0.1cm} \\
Ti &$s$& -0.22369 & -0.22357 & -0.5 & -0.0 &   1.5 &  10.7 &  11.7 \vspace{ 0.0cm} \\
   &$d$& -0.39560 & -0.39542 &  5.3 &  0.1 &   6.0 &   6.1 &  17.6 \vspace{+0.1cm} \\
Cr &$s$& -0.21104 & -0.21127 & -1.4 &  0.1 & -13.5 &  -7.9 & -22.7 \vspace{ 0.0cm} \\
   &$d$& -0.31518 & -0.31494 &  2.8 &  0.6 &  17.4 &   2.8 &  23.6 \vspace{+0.1cm} \\
Fe &$s$& -0.26375 & -0.26375 & -1.5 &  0.1 &   1.1 &   0.7 &   0.4 \vspace{ 0.0cm} \\
   &$d$& -0.59311 & -0.59295 &  8.7 &  1.1 &   3.3 &   3.1 &  16.3 \vspace{+0.1cm} \\
Br &$s$& -1.02099 & -1.02101 & -0.8 & -1.5 &   0.0 &   0.3 &  -1.9 \vspace{ 0.0cm} \\
   &$p$& -0.45616 & -0.45600 & -0.1 & 16.2 &   0.0 &  -0.1 &  16.0 \vspace{+0.1cm} \\
Mo &$s$& -0.21217 & -0.21270 & -0.5 &  1.2 & -41.2 & -11.9 & -52.4 \vspace{ 0.0cm} \\
   &$d$& -0.29622 & -0.29585 &  0.5 &  5.5 &  36.8 &  -5.5 &  37.3 \vspace{+0.1cm} \\
Ag &$s$& -0.23716 & -0.23712 & -1.0 &  3.9 &   0.0 &   1.6 &   4.5 \vspace{ 0.0cm} \\
   &$d$& -0.51128 & -0.51108 &  0.8 & 21.1 &   0.0 &  -2.8 &  19.2 \vspace{+0.1cm} \\
Sb &$s$& -0.62683 & -0.62689 & -0.3 & -1.2 &  -5.3 &   0.5 &  -6.4 \vspace{ 0.0cm} \\
   &$p$& -0.30062 & -0.29996 &  0.0 & 54.7 &  13.4 &  -2.3 &  65.8 \vspace{+0.1cm} \\
\hline
\multicolumn{1}{l}{}&
\multicolumn{3}{c}{$s$-block}&  0.1 &  0.0 & 0.0 &  0.2 &  0.3 \\
\multicolumn{1}{l}{Mean $|\Delta \epsilon|$}&
\multicolumn{3}{c}{$p$-block}&  0.8 &  9.4 & 1.2 &  0.6 & 11.1 \\
\multicolumn{1}{l}{}&
\multicolumn{3}{c}{$d$-block}&  1.9 & 19.7 & 7.4 &  7.3 & 29.0 \\
\hline \hline
\end{tabular}
\caption{Eigenvalues (a.u.) for the neutral ground states of selected
atoms.  Eigenvalues obtained from HF pseudopotential calculations with
$jj$ coupling are compared with AE DF eigenvalues. All eigenvalues
shown are $j$-weighted averages.
\newline
Errors are due to: \newline
$\Delta\epsilon_1$ - transforming the Dirac equation to a Schr\"odinger equation in the valence region.
\newline
$\Delta\epsilon_2$ - higher order effects due to excluding SO coupling.
\newline
$\Delta\epsilon_3$ - localization.
\newline
$\Delta\epsilon_4$ - parameterisation.
\newline
$\Delta\epsilon_{tot}=\epsilon^{jj}_l-\epsilon^{AE}_l=\sum_i \Delta\epsilon_i$ is the total error.}
\label{tab:1}
\end{table}

\begin{table}[t]
\begin{tabular}{lcccc}
\hline \hline
\multicolumn{1}{c}{Atom}&\multicolumn{1}{c}{Orbital}&
\multicolumn{1}{c}{AE DF}&
\multicolumn{1}{c}{Our}  & 
\multicolumn{1}{c}{PC} \\
\hline
Li &$s$& -0.19634 & -0.19634 & -0.19646  \vspace{+0.1cm} \\
Be &$s$& -0.30932 & -0.30933 & -0.30955  \vspace{+0.1cm} \\
C  &$s$& -0.71212 & -0.71213 & -0.71406  \vspace{ 0.0cm} \\
   &$p$& -0.40867 & -0.40869 & -0.40916  \vspace{+0.1cm} \\
Si &$s$& -0.54522 & -0.54522 & -0.54653  \vspace{ 0.0cm} \\
   &$p$& -0.27967 & -0.27968 & -0.27885  \vspace{+0.1cm} \\
Ti &$s$& -0.22369 & -0.22357 & -0.22365  \vspace{ 0.0cm} \\
   &$d$& -0.39560 & -0.39542 & -0.39546  \vspace{+0.1cm} \\
Cr &$s$& -0.21104 & -0.21127 & -0.21109  \vspace{ 0.0cm} \\
   &$d$& -0.31518 & -0.31494 & -0.31513  \vspace{+0.1cm} \\
Fe &$s$& -0.26375 & -0.26375 & -0.26374  \vspace{ 0.0cm} \\
   &$d$& -0.59311 & -0.59295 & -0.59314  \vspace{+0.1cm} \\
Br &$s$& -1.02099 & -1.02101 & -1.02111  \vspace{ 0.0cm} \\
   &$p$& -0.45616 & -0.45600 & -0.45660  \vspace{+0.1cm} \\
Mo &$s$& -0.21217 & -0.21270 & -0.21602  \vspace{ 0.0cm} \\
   &$d$& -0.29622 & -0.29585 & -0.28839  \vspace{+0.1cm} \\
Ag &$s$& -0.23716 & -0.23712 & -0.23623  \vspace{ 0.0cm} \\
   &$d$& -0.51128 & -0.51108 & -0.51140  \vspace{+0.1cm} \\
Sb &$s$& -0.62683 & -0.62689 & -0.62690  \vspace{ 0.0cm} \\
   &$p$& -0.30062 & -0.29996 & -0.29950  \vspace{+0.1cm} \\
\hline
\multicolumn{3}{l}{Mean error}          & 0.00006 & 0.00011 \\
\multicolumn{3}{l}{Mean absolute error} & 0.00015 & 0.00098 \\
\hline \hline
\end{tabular}
\caption{Eigenvalues (a.u.) for the neutral ground states of selected
atoms.  The eigenvalues obtained from AE DF calculations are compared
with those from HF pseudopotential calculations with $jj$ coupling
using our soft parameterized pseudopotentials, and with those from the
singular parameterized pseudopotentials of Pacios and
Christiansen~\cite{christiansen79,christiansen_web} (PC).  All
eigenvalues shown are $j$-weighted averages.}
\label{tab:2}
\end{table}

\begin{table}[t]
\center{\begin{tabular}{llccc}
\hline \hline
\multicolumn{2}{c}{ }                      &
\multicolumn{1}{c}{AE HF}                  &
\multicolumn{2}{c}{ }                      \\
\multicolumn{1}{c}{Atom}&
\multicolumn{1}{c}{Configuration}&
\multicolumn{1}{c}{with RC}&
\multicolumn{1}{c}{Our }&
\multicolumn{1}{c}{PC}  \\
\hline
Li&$2s^{1}            [^2S]$& 0.00000 & 0.00000 & 0.00000 \vspace{ 0.0cm} \\
  &$2p^{1}            [^2P]$& 0.06767 & 0.06769 & 0.06782 \vspace{ 0.0cm} \\
  &$3d^{1}            [^2D]$& 0.14076 & 0.14078 & 0.14089 \vspace{+0.1cm} \\
Be&$2s^{2}            [^1S]$& 0.00000 & 0.00000 & 0.00000 \vspace{ 0.0cm} \\
  &$2s^{1}2p^{1}      [^3P]$& 0.06157 & 0.05951 & 0.06076 \vspace{ 0.0cm} \\
  &$2s^{1}3d^{1}      [^3D]$& 0.23892 & 0.23910 & 0.23942 \vspace{+0.1cm} \\
C &$2s^{2}2p^{2}      [^3P]$& 0.00000 & 0.00000 & 0.00000 \vspace{ 0.0cm} \\
  &$2s^{1}2p^{3}      [^5S]$& 0.08982 & 0.08461 & 0.08663 \vspace{ 0.0cm} \\
  &$2s^{2}2p^{1}3d^{1}[^3F]$& 0.34003 & 0.34060 & 0.34102 \vspace{+0.1cm} \\
Si&$3s^{2}3p^{2}      [^3P]$& 0.00000 & 0.00000 & 0.00000 \vspace{ 0.0cm} \\
  &$3s^{1}3p^{3}      [^5S]$& 0.09313 & 0.09076 & 0.09243 \vspace{ 0.0cm} \\
  &$3s^{2}3p^{1}3d^{1}[^3F]$& 0.21425 & 0.21458 & 0.21390 \vspace{+0.1cm} \\
Ti&$3d^{2}4s^{2}      [^3F]$& 0.00000 & 0.00000 & 0.00000 \vspace{ 0.0cm} \\
  &$3d^{3}4s^{1}      [^5F]$& 0.02505 & 0.02945 & 0.02957 \vspace{ 0.0cm} \\
  &$3d^{4}            [^5D]$& 0.16459 & 0.18113 & 0.18015 \vspace{+0.1cm} \\
Cr&$3d^{5}4s^{1}      [^7S]$& 0.00000 & 0.00000 & 0.00000 \vspace{ 0.0cm} \\
  &$3d^{4}4s^{2}      [^5D]$& 0.03914 & 0.04882 & 0.04820 \vspace{ 0.0cm} \\
  &$3d^{6}            [^5D]$& 0.26201 & 0.27226 & 0.26986 \vspace{+0.1cm} \\
Fe&$3d^{6}4s^{2}      [^5D]$& 0.00000 & 0.00000 & 0.00000 \vspace{ 0.0cm} \\
  &$3d^{7}4s^{1}      [^5F]$& 0.07555 & 0.07984 & 0.08006 \vspace{ 0.0cm} \\
  &$3d^{8}            [^3F]$& 0.28933 & 0.30548 & 0.30477 \vspace{+0.1cm} \\
Br&$4s^{2}4p^{5}      [^2P]$& 0.00000 & 0.00000 & 0.00000 \vspace{ 0.0cm} \\
  &$4s^{1}4p^{6}      [^2S]$& 0.55464 & 0.55820 & 0.55803 \vspace{ 0.0cm} \\
  &$4s^{2}4p^{4}4d^{1}[^4F]$& 0.33407 & 0.33376 & 0.33440 \vspace{+0.1cm} \\
Mo&$4d^{5}5s^{1}      [^7S]$& 0.00000 & 0.00000 & 0.00000 \vspace{ 0.0cm} \\
  &$4d^{4}5s^{2}      [^5D]$& 0.08763 & 0.10114 & 0.09298 \vspace{ 0.0cm} \\
  &$4d^{6}            [^5D]$& 0.15962 & 0.17646 & 0.18216 \vspace{+0.1cm} \\
Ag&$4d^{10}5s^{1}     [^2S]$& 0.00000 & 0.00000 & 0.00000 \vspace{ 0.0cm} \\
  &$4d^{9}5s^{2}      [^2D]$& 0.14411 & 0.14888 & 0.15060 \vspace{ 0.0cm} \\
  &$4d^{10}5p^{1}     [^2P]$& 0.11066 & 0.10873 & 0.11195 \vspace{+0.1cm} \\
Sb&$5s^{2}5p^{3}      [^4S]$& 0.00000 & 0.00000 & 0.00000 \vspace{ 0.0cm} \\
  &$5s^{1}5p^{4}      [^4P]$& 0.27820 & 0.28326 & 0.28370 \vspace{ 0.0cm} \\
  &$5p^{5}            [^2P]$& 0.76145 & 0.78007 & 0.78147 \vspace{+0.1cm} \\
\hline
\multicolumn{3}{l}{Mean error}          &  0.00513 & 0.00539           \\
\multicolumn{3}{l}{Mean absolute error} &  0.00621 & 0.00585           \\
\hline \hline
\end{tabular}}
\caption{Excitation energies (a.u.) for selected atoms.  Excitation
energies obtained from AE HF calculations with relativistic
corrections (RC), are compared with those from HF pseudopotential
calculations with $LS$ coupling.  Results are shown for both our soft
parameterized pseudopotentials, and for the singular parameterized
pseudopotentials of Pacios and
Christiansen~\cite{christiansen79,christiansen_web} (PC). }
\label{tab:3}
\end{table}

\begin{table}[t]
\begin{tabular}{lccc} \hline \hline
Atom\ \ \ \ &
\ \ \ \ $E_{ion}^{AE}$\ \ \ \ &
\ \ \ \ $\tilde{E}_{ion}$ \ \ \ \ &
\ \ \ \ $E_{ion}^{PC}$ \\ \hline
Li & 0.19632 & 0.19634 & 0.19646 \\
Be & 0.29566 & 0.29584 & 0.29618 \\
C  & 0.39625 & 0.40032 & 0.39724 \\
Si & 0.28083 & 0.28743 & 0.28083 \\
Ti & 0.20374 & 0.20044 & 0.20115 \\
Cr & 0.22054 & 0.22849 & 0.22757 \\
Fe & 0.23288 & 0.22623 & 0.22792 \\
Br & 0.39481 & 0.39473 & 0.39530 \\
Mo & 0.22801 & 0.24066 & 0.24459 \\
Ag & 0.23088 & 0.23353 & 0.23274 \\
Sb & 0.31787 & 0.31923 & 0.31852 \\
\hline
\multicolumn{2}{l}{Mean error}          & 0.00231 & 0.00188 \\
\multicolumn{2}{l}{Mean absolute error} & 0.00414 & 0.00325 \\
\hline \hline
\end{tabular}
\caption{Comparison of first ionization potentials obtained from AE
calculations ($E_{ion}^{AE}$), from HF calculations using our soft
parameterized pseudopotentials ($\tilde{E}_{ion}$), and for the
singular parameterized pseudopotentials of Pacios and
Christiansen~\cite{christiansen79,christiansen_web} ($E_{ion}^{PC}$)
($LS$ coupling is used throughout).}
\label{tab:4}
\end{table}

\begin{table}[t]
\begin{tabular}{lcddd} \hline \hline
\multicolumn{1}{c}{Atom} &
\multicolumn{1}{c}{Configuration} &
\multicolumn{1}{c}{AE DF}&
\multicolumn{1}{c}{Our}  &
\multicolumn{1}{c}{PC}   \\
\hline
Li        &$p^1$&    0.6 &     0.8 &     0.9 \\
Be$^{+}$  &$p^1$&    8.9 &     8.5 &     9.3 \\
B         &$p^1$&   20.6 &    20.0 &    21.8 \\
C$^{+}$   &$p^1$&   79.3 &    75.8 &    84.5 \\
N$^{2+}$  &$p^1$&  209.8 &   197.9 &   219.9 \\
O$^{3+}$  &$p^1$&  452.3 &   454.4 &   498.9 \\
F         &$p^5$&  410.5 &   402.4 &   438.9 \\
Ne$^{+}$  &$p^5$&  815.4 &   820.5 &   892.0 \\
Al        &$p^1$&  120.0 &   116.1 &   114.8 \\
Si$^{+}$  &$p^1$&  308.7 &   293.9 &   292.2 \\
P$^{2+}$  &$p^1$&  599.3 &   556.1 &   553.9 \\
S$^{3+}$  &$p^1$& 1019.4 &  1020.0 &  1013.5 \\
Cl        &$p^5$&  920.6 &   903.9 &   910.3 \\
Ar$^{+}$  &$p^5$& 1494.1 &  1472.6 &  1487.4 \\
Sc        &$d^1$&  165.7 &   169.2 &   191.0 \\
Ti$^{+}$  &$d^1$&  370.1 &   410.7 &   458.4 \\
Ti$^{3+}$ &$d^1$&  405.4 &   440.6 &   487.3 \\
Cu        &$d^9$& 2122.9 &  2110.5 &  2292.9 \\
Zn$^{+}$  &$d^9$& 2820.0 &  2814.1 &  3036.3 \\
Ga        &$p^1$&  811.8 &   756.6 &   772.0 \\
Ge$^{+}$  &$p^1$& 1764.0 &  1681.9 &  1710.4 \\
Br        &$p^5$& 3744.5 &  3777.9 &  3867.3 \\
Kr$^{+}$  &$p^5$& 5458.8 &  5944.1 &  5431.0 \\
\hline
\multicolumn{3}{l}{Mean $\%$ error}          &  0.4 & 5.7 \\
\multicolumn{3}{l}{Mean absolute $\%$ error} &  4.8 & 8.2 \\
\hline \hline
\end{tabular}
\caption{ Spin-orbit splitting (cm$^{-1}$) for selected atoms.
Splitting obtained from all-electron DF calculations (with no
relativistic correction) are compared with those resulting from
$jj$-coupled HF calculations using parameterized AREP and SO
pseudopotentials.  Results are given for both our soft parameterized
AREP and SO pseudopotentials, and those of Christiansen
\textit{et al.}~\cite{christiansen79,ross86b,christiansen_web}
(PC).}
\label{tab:5}
\end{table}

\end{document}